\documentclass{article}

\usepackage{arxiv}

\usepackage[utf8]{inputenc} 
\usepackage[T1]{fontenc}    
\usepackage{hyperref}       
\usepackage{url}            
\usepackage{booktabs}       
\usepackage{amsfonts}       
\usepackage{amssymb}        
\usepackage{amsmath}        
\usepackage{nicefrac}       
\usepackage{microtype}      
\usepackage{lipsum}		
\usepackage{graphicx}
\usepackage[numbers]{natbib}
\usepackage{doi}
\usepackage{wrapfig}

\title{Introducing the vfunc R package}

\author{ \href{https://orcid.org/0000-0001-5982-0415}{\includegraphics[width=0.03\textwidth]{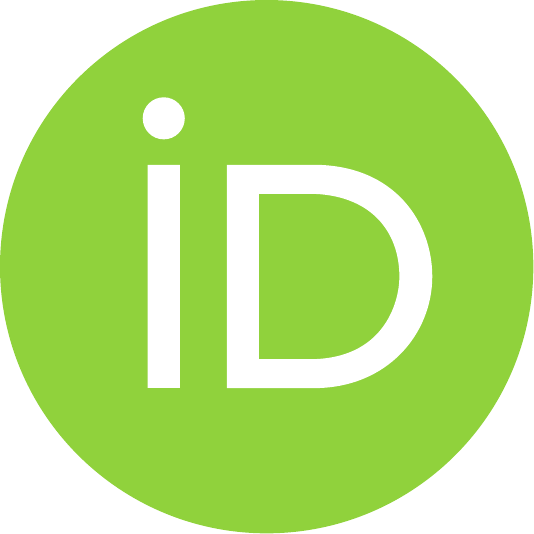}\hspace{1mm}Robin K. S.~Hankin}\thanks{\href{https://academics.aut.ac.nz/robin.hankin}{work};  
\href{https://www.youtube.com/watch?v=JzCX3FqDIOc&list=PL9_n3Tqzq9iWtgD8POJFdnVUCZ_zw6OiB&ab_channel=TrinTragulaGeneralRelativity}{play}} \\
 University of Stirling\\
	\texttt{hankin.robin@gmail.com} \\
}

\hypersetup{
pdftitle={The vfunc R package},
pdfsubject={q-bio.NC, q-bio.QM},
pdfauthor={Robin K. S.~Hankin},
pdfkeywords={functional programming}
}

\usepackage{Sweave}
\begin{document}
\maketitle

\begin{abstract}
In mathematics, given two functions
$f,g\colon\mathbb{R}\longrightarrow\mathbb{R}$, it is natural to
define $f+g$ as the function that maps $x\in\mathbb{R}$ to $f(x) +
g(x)$.  However, in base R, objects of class {\tt function} do not
have arithmetic methods defined, so idiom such as ``{\tt f + g}''
returns an error, even though it has a perfectly reasonable
expectation.  The {\tt vfunc package} offers this functionality.
Other similar features are provided, which lead to compact and
readable idiom.  A wide class of coding bugs is eliminated.
\end{abstract}

\section{Introduction}

\setlength{\intextsep}{0pt}
\begin{wrapfigure}[3]{r}{0.2\textwidth}
  \begin{center}
    \includegraphics[width=1in]{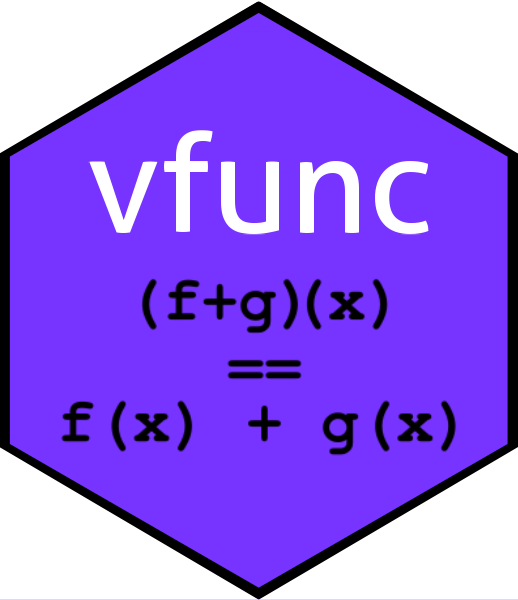}
  \end{center}
\end{wrapfigure}

Consider the following R session:
\begin{Schunk}
\begin{Sinput}
> f <- function(x){x^2}
> g <- function(x){1/(1-x)}
> f + g
\end{Sinput}
\begin{Soutput}
Error in f + g : non-numeric argument to binary operator
\end{Soutput}
\end{Schunk}

Above, there is a reasonably clear expectation for ``{\tt f + g}'': it
should give a function that returns the sum of {\tt f()} and {\tt
  g()}; something like {\tt function(x)\{f(x) + g(x)\}}.  However, it
returns an error because {\tt f} and {\tt g} are objects of {\tt S4}
class {\tt function}, which do not have an addition method.  Further,
it is not possible to define {\tt Arith} group {\tt S4} methods [in
  this case, overloading addition, ``{\tt +}''] so that this idiom
operates as desired.  This is because the {\tt function} class is {\em
  sealed} in {\tt S4}: the definition of new methods for it is
prohibited.  Here I present the {\tt vfunc} R~\cite{rcore2024} package
that furnishes appropriate idiom.  It is available on CRAN at
\url{https://CRAN.R-project.org/package=vfunc}.  The package defines a
new {\tt S4} class {\tt vf} (``virtual function'') which inherits from
{\tt function}, but for which new methods can be defined.  This device
furnishes some ways to apply {\tt Arith} methods for functions.  

\section{The package in use}

The package is designed so that objects of class {\tt vf} operate as
functions but are subject to arithmetic operations, which are executed
transparently.  For example:

\begin{Schunk}
\begin{Sinput}
> library("vfunc")
> f <- as.vf(f)
> g <- as.vf(g)
> (f + g)(1:10)
\end{Sinput}
\begin{Soutput}
 [1]      Inf  3.00000  8.50000 15.66667 24.75000 35.80000 48.83333 63.85714
 [9] 80.87500 99.88889
\end{Soutput}
\end{Schunk}

Above, we coerce {\tt f} and {\tt g} to objects of {\tt S4} class {\tt vf} [for
"virtual function"].  Such objects have {\tt Arith} methods defined and
may be combined arithmetically; for example addition is dispatched to

\begin{Schunk}
\begin{Soutput}
function(e1, e2){as.vf(function(...){e1(...) + e2(...)})}
\end{Soutput}
\end{Schunk}

The {\tt vf} class has a single {\tt .Data} slot of type {\tt
"function"} which means that objects of this class inherit much of the
behaviour of base class {\tt function}; above, we see that {\tt e1}
and {\tt e2} may be executed with their argument list directly.  In
practice this means that {\tt f+g} behaves as intended, and suggests
other ways in which it can be used:

\begin{Schunk}
\begin{Sinput}
> (f + 4*g - f*g)(1:10)
\end{Sinput}
\begin{Soutput}
 [1]       NaN   4.00000  11.50000  20.00000  30.25000  42.40000  56.50000
 [8]  72.57143  90.62500 110.66667
\end{Soutput}
\end{Schunk}

The advantages of such idiom fall in to two main categories.  Firstly,
code can become considerably more compact; and secondly one can guard
against a wide class of hard-to-find bugs.  Now consider {\tt f()} and
{\tt g()} to be trivariate functions, each taking three arguments,
say,

\begin{Schunk}
\begin{Sinput}
> f <- function(x,y,z){x + x*y - x/z}
> g <- function(x,y,z){x^2 - z}
\end{Sinput}
\end{Schunk}

and $x=1.2$, $y=1.7$, $z=4.3$.  Given this, we wish to calculate

$$(f(x,y,z) + g(x,y,z))(f(x,y,z) + 4 - 2f(x,y,z)g(x,y,z)).$$

How would one code up such an expression in R?  The standard way would be

\begin{Schunk}
\begin{Sinput}
>  x <- 1.2
>  y <- 1.7
>  z <- 4.3		
> (f(x,y,z) + g(x,y,z))*(f(x,y,z) + 4 - 2*f(x,y,z)*g(x,y,z))
\end{Sinput}
\begin{Soutput}
[1] 2.411975
\end{Soutput}
\end{Schunk}

Note the repeated specification of argument list {\tt (x,y,z)}, repeated
here five times.  Now use the {\tt vfunc} package:

\begin{Schunk}
\begin{Sinput}
> f <- as.vf(f)
> g <- as.vf(g)
> ((f + g)*(f + 4 - 2*f*g))(x,y,z)
\end{Sinput}
\begin{Soutput}
[1] 2.411975
\end{Soutput}
\end{Schunk}

See how the package allows one to ``factorize'' the argument list so
it appears once, leading to more compact code.  It is also arguably
less error-prone, as the following example illustrates.  Consider

$$
f(x+z,y+z,f(x,x,y)-g(x,x,y)) + g(x+z, y+z,f(x,x,y)-g(x,x,y))
$$

(such expressions arise in the study of dynamical systems).  Note that
functions $f$ and $g$ are to be evaluated with two distinct sets of
arguments at different levels of nesting, namely $(x,x,y)$ at the
inner level and $(x+z,y+z,f(x,x,y)-g(x,x,y)$ at the outer.  Standard R
idiom would be

\begin{Schunk}
\begin{Sinput}
> f(x + z, y + z, f(x, x, y) - g(x, x, y)) + g(x + z, y + z, f(x, x, y) - g(x, x, y))
\end{Sinput}
\begin{Soutput}
[1] 64.04918
\end{Soutput}
\end{Schunk}

The author can attest that finding bugs in such expressions can be
difficult [it is easy to mistype {\tt (x,x,y)} in one of its
occurences, yet difficult to detect the error].  However, {\tt vfunc}
idiom would be

\begin{Schunk}
\begin{Sinput}
> (f + g)(x + z, y + z, (f - g)(x, x, y))
\end{Sinput}
\begin{Soutput}
[1] 64.04918
\end{Soutput}
\end{Schunk}

which is certainly shorter, arguably neater and at least the author
finds such constructions considerably less error-prone.  In this form,
one can be sure that both {\tt f()} and {\tt g()} are called with
identical arguments at each of the two levels in the expression, as
the arguments appear only once.

\subsection{Overloading}

Looking again at the method for {\tt vf} addition, viz

\begin{Schunk}
\begin{Soutput}
function(e1, e2){as.vf(function(...){e1(...) + e2(...)})}
\end{Soutput}
\end{Schunk}

we see the {\tt +} operator is used to sum the return values of {\tt
  e1()} and {\tt e2()}.  There is no reason that this operator cannot
itself be overloaded, and the {\tt vfunc} package works transparently
if this is the case, with either {\tt S3} or {\tt S4}.  Taking the
{\tt onion} package~\cite{hankin2006_onion} as an example:

\begin{Schunk}
\begin{Sinput}
> library("onion")
> options("show_onions_compactly" = TRUE)
> f <- as.vf(function(x,y){x + x*y})
> g <- as.vf(function(x,y){x^2 + y})
> (f + g - f*g)(1 + Hj,Hk)
\end{Sinput}
\begin{Soutput}
        Re 
4+2i+2j-1k 
\end{Soutput}
\end{Schunk}

\section{Primitive functions}

The R language includes a number of primitive functions as {\tt S4}
Math generics, including the trig functions such as {\tt sin()}, and a
few others such as the cumulative sum {\tt cumsum()}.  These functions
are quite deep-seated and cannot easily be modified to work with
objects of class {\tt vf}.  The package defines capitalized versions
of primitive functions to operate with other objects of class {\tt
vf}.  Taking {\tt sin()} as an example we have

\begin{Schunk}
\begin{Sinput}
> vfunc::Sin
\end{Sinput}
\begin{Soutput}
An object of class "vf"
function (x) 
{
    sin(x)
}
<bytecode: 0x5b9452ac73c0>
<environment: namespace:vfunc>
\end{Soutput}
\end{Schunk}

Then we may, for example, combine trig functions with user-defined functions:

\begin{Schunk}
\begin{Sinput}
> fun <- as.vf(function(x){x^2 + 2})
> (fun(Sin) + Sin(fun) - 3*Sin*fun)(0.32)
\end{Sinput}
\begin{Soutput}
[1] 0.9769132
\end{Soutput}
\end{Schunk}

Above, we see package idiom being used to evaluate $\sin^2(0.32) + 3 +
\sin(0.32^2+2) - 3\cdot\sin 0.32\cdot(0.32^2+2)$.  In base R:

\begin{Schunk}
\begin{Sinput}
> fun(sin(0.32)) + sin(fun(0.32)) - 3*sin(0.32)*fun(0.32)
\end{Sinput}
\begin{Soutput}
[1] 0.9769132
\end{Soutput}
\end{Schunk}

This construction allows one to define composite functions such as

\begin{Schunk}
\begin{Sinput}
> j <- as.vf(function(x,y){Cos(x) + Sin(x-y)})
> k <- as.vf(function(x,y){Tan(x) + Log(x+y)})
> l <- as.vf(function(x,y){Sin(x/2) + x^2   })
\end{Sinput}
\end{Schunk}

(note that functions {\tt j()}, {\tt k()} and {\tt l()} are
bivariate).  Then compare

\begin{Schunk}
\begin{Sinput}
> (j + k + l)(Sin + Log, Cos + Exp)(Sin + Tan)(0.4)
\end{Sinput}
\begin{Soutput}
[1] 2.545235
\end{Soutput}
\end{Schunk}

with the one-stage idiom which reads:

\begin{Schunk}
\begin{Soutput}
[1] 2.545235
\end{Soutput}
\end{Schunk}

\begin{Schunk}
\begin{Sinput}
> j(sin(sin(0.4) + tan(0.4)) + log(sin(0.4) + tan(0.4)), cos(sin(0.4) + tan(0.4)) +
+ exp(sin(0.4) + tan(0.4))) + k(sin(sin(0.4) + tan(0.4)) + log(sin(0.4) + tan(0.4)),
+ cos(sin(0.4) + tan(0.4)) + exp(sin(0.4) + tan(0.4)))+ l(sin(sin(0.4) + tan(0.4)) +
+ log(sin(0.4) + tan(0.4)), cos(sin(0.4) + tan(0.4)) + exp(sin(0.4) + tan(0.4)))
\end{Sinput}
\begin{Soutput}
[1] 2.545235
\end{Soutput}
\end{Schunk}

and the multi-stage idiom:

\begin{Schunk}
\begin{Sinput}
> A <- function(x,y){j(x,y) + k(x,y) + l(x,y)}
> B <- function(x){sin(x) + log(x)}
> C <- function(x){cos(x) + exp(x)}
> D <- function(x){sin(x) + tan(x)}
> x <- 0.4
> A(B(D(x)), C(D(x)))
\end{Sinput}
\begin{Soutput}
[1] 2.545235
\end{Soutput}
\end{Schunk}

See how the one-stage idiom is very long, and the multi-stage idiom is
opaque [and nevertheless has repeated instances of {\tt (x,y)} and
  {\tt x}].

\section{Conclusions}

The {\tt vfunc} package allows functions to be ``factorized'', that
is, {\tt f(x) + g(x)} to be re-written {\tt(f + g)(x)}.  This allows
for concise idiom and eliminates a certain class of coding errors.
The package also allows for recursive application of such ideas.
Further work might include an assessment of the package's
computational efficiency.

\bibliographystyle{apalike}
\bibliography{vfunc}

\end{document}